\documentstyle[preprint,aps]{revtex}

\begin{document}

\title
{Renormalization - group analysis of dilute Bose system in $d$ - dimension at finite temperature}

\author
{M. Crisan, D. Bodea, I. Grosu, and I. Tifrea\cite{clem}}

\address
{Department of Theoretical Physics, University of Cluj, 3400 Cluj-Napoca, Romania}

\maketitle

\begin{abstract}
We study the $d$ - dimensional Bose gas at finite temperature using the renormalization group
method. The flow - equations and the free energy have been obtained for dimension $d$, and the
cases $d<2$ and $d=2$ have been analysed in the limit of low and high temperatures. The critical
temperature, the coherence length and the specific heat of a two dimensional Bose gas have been
obtained using a solution for the coupling constant which does not present a singular behavior. 
\end{abstract}

\newpage
\narrowtext

\section{Introduction}

Renewed interest for two dimensional dilute Bose gas has been recently determined by the 
discovery of high - temperature superconductors.

The Bose - Einstein condensation is known not to occur at finite temperature for the interacting
boson system in one and two dimensions (the Mermin - Wagner theorem); the absence of a 
condensate does not necessary imply the absence of a phase transition (in the universality class 
as $XY$ model) and it is expected to be in the superfluid state.

In the last years the study of two dimensional ($2D$) Bose condensation was stimulated by new 
experimental results concerning scattering in dilute gases of alkaline atoms, spin polarized 
hydrogen recombining and adsorbed helium monoalloys \cite{1}.

The theory of weekly interacting Bose gas have been developed first at $T=0$ by Bogolibov \cite{2}
and Lee, Huang and Yang \cite{3} at finite temperature. Using the $T=0$ results Hohenberg and 
Martin \cite{4}, Popov and Fadeev \cite{5}, Singh \cite{6} and Cheung and Griffin \cite{7} 
developed the field - theoretical methods for finite temperature. Using this method it was
possible to calculate the observable physical quantities, but these calculations are quite
difficult because of the influence of correlations induced by the condensed phase.

The many - body theory developed by Popov \cite{8,9} is manly based on the $t$ - matrix 
approximation, also applied for the $2D$ Bose gas by Schick \cite{11}. We mention that the 
$t$ - matrix for the scattering in the Bose system involves a sum over all ladder diagrams to 
infinite order in interaction, taking the repeated scattering of two particles.

The purpose of this paper is the study of the critical region of $2D$ interacting Bose gas, 
which was first done in the famous paper of Fisher and Hohenberg \cite{12}, using $t$ - matrix
method and the Renormalization Group (RG) method. The problem was also studied by Kolomeisky and
Straley \cite{13} for a $D$ dimensional Bose system at $T=0$. The influence of a gauge - field 
on a superfluid phase transition in a two dimensional model at finite temperature has been 
studied by Ubbens et al \cite{14} and for the case of three dimensional dilute Bose gas by
Biglsma and Stoof \cite{5}.

As we mentioned, the results of RG study from Ref. \cite{12}
have been used for the problem of the quantum phase transitions in the spin systems 
\cite{16}, or proposed \cite{17} to study the large scale behavior in the crossover
problem from high temperature superconductivity. All these papers used the approximations from
Ref. \cite{12}, which will be reconsidered especially concerning the solutions of the RG scaling
equations. We will also recalculate the number of condensate bosons for $d=2$ and the coherence
length using the new result obtained by solving more accurate the equations for the coupling 
constants.

The paper is organized as follows. In Sec. II we present the model and write down the RG scaling
equations. In Sec. III we solve the equations in the low temperature region and calculate the 
number of condensate bosons and the coherence length $\xi(T)$ for the general case of dimension
$d$. The interesting case $d=2$ will be treated in Sec. IV where we also calculate the 
temperature dependence of the specific heat. The high temperature behavior will be analyzed in 
Sec. V. Sec. VI contain the relevance of our results compared with the results from literature.

\section{Model and Scaling Equations}

We consider the dilute Bose system in $d$ dimensions at finite temperature $T$ described by the
action:
\begin{equation}
S_{eff} = S_{eff}^{(2)} + S_{eff}^{(4)}
\label{2.1}
\end{equation}
where
\begin{equation}
S_{eff}^{(2)} = \frac{1}{2} \sum_k \left[ \frac{\hbar^2 k^2}{2 m} - \mu 
- \frac{|\omega_n|}{\Gamma} \right] |\phi(k)|^2
\label{2.2}
\end{equation}
and
\begin{equation}
S_{eff}^{(4)}  = \frac{u}{4} \sum_{k_1} \ldots \sum_{k_4} \phi(k_1) \cdots \phi(k_4)
\delta \left( \sum_{i=1}^4 k_i \right)
\label{2.3}
\end{equation}
and the following notation have been used:
\begin{equation}
\sum_k \cdots \rightarrow k_B T \sum_n \int \frac{d^d {\bf k}}{(2 \pi)^d} \cdots
\label{2.4}
\end{equation}
$\omega_n = 2 \pi n k_B T$ being the bosonic frequencies.

In Eq. (\ref{2.2}) $\mu$ is the chemical potential of the bosonic system, described by the scalar
field $\phi(k)$ and $\Gamma$ is an energy parameter which controls the strength of the quantum
fluctuations, the classical limit being $\Gamma = 0$. The Eq. (\ref{2.3}) represent the 
interaction between fluctuations, $u$ ($u>0$) being the coupling constant.

The basic propagator of the bosonic system has the form:
\begin{equation}
G({\bf k}, \omega_n) = \left[ \frac{\hbar^2 k^2}{2 m} - \mu - \frac{|\omega_n|}{\Gamma} \right]^{-1}
\label{2.5}
\end{equation}
and the renormalization transformations are carried out by integrating over a momentum shell and
summing over all frequencies.

Regarding this procedure we mention that the summation over the bosonic frequencies (transformed
in integral) has to be done for all frequencies and not on a shell as in the method applied in
\cite{18,19} for bosonic excitations. The physical argument is that in this case the presence of 
the chemical potential is important because we expect different contributions on the energy 
scale, which is in fact identical with a change of physical behavior in different temperatures
domain. On the other hand, the limit cases $\mu = \pm \infty$ correspond to the condensate, 
respectively normal states. 

After integrating out the degrees of freedom in the momentum shell, we rescale the variables as:
\begin{equation}
k = \frac{k'}{b} \; , \; \omega_n = \frac{{\omega_n}'}{b^z} \; , \; T = \frac{T'}{b^z}
\label{2.6}
\end{equation}
and the field operator as:
\begin{equation}
\phi'(k',{\omega_n}') = b^{-(d+z+2)/2} \phi \left( \frac{k'}{b}, \frac{{\omega_n}'}{b^z} \right)
\label{2.7}
\end{equation}
where $z$ is the dynamic critical coefficient and we will introduce for $b$ the parameterization
$b  = e^l$ ($b>0 \; , \; l>0$). The scaling equations obtained in Ref. \cite{12,14} for 
$T \ne 0$ and in \cite{13} for $T=0$ will be written as:
\begin{equation}
\frac{d \Gamma(l)}{d l} = - (2-z) \Gamma(l)
\label{2.8}
\end{equation}
\begin{equation}
\frac{d T(l)}{d l} = z T(l)
\label{2.9}
\end{equation}
\begin{equation}
\frac{d \mu(l)}{d l} = 2 \mu(l) - K_d F_{\mu} [\mu(l),T(l),u(l),\Gamma(l)]
\label{2.10}
\end{equation}
\begin{eqnarray}
\frac{d u(l)}{d l} & = & [4-(d+z)] u(l) - \frac{1}{4} K_d 
\left\{ 8 F_{p-p}[\mu(l),T(l),u(l),\Gamma(l)] \right. \nonumber \\
& & \left. + 2 F_{p-a} [\mu(l),T(l),u(l),\Gamma(l)] \right\} u^2(l)
\label{2.11}
\end{eqnarray}
\begin{equation}
\frac{d F(l)}{d l} = (d+z) F(l) + K_d F_f [\mu(l),T(l),u(l),\Gamma(l)]
\label{2.12}
\end{equation}
where $F(l)$ is the free energy and $K_d = \pi^{d/2} / \left( 2^{d-1} \Gamma(d/2) \right)$. The
constants $F_{\mu}$, $F_{p-p}$, $F_{p-a}$ and $F_f$ from Eqs. (\ref{2.10} - \ref{2.12}) are 
given by the expressions:
\begin{eqnarray}
F_{\mu} & = & F_{\mu}[\mu(l),T(l),\Gamma(l)] \nonumber \\ 
& = & 2 \mu(l) - 
\frac{K_d \Lambda^2 \Gamma(l)}{\exp \left[ \frac{\Gamma(l)}{k_b T} 
\left( \frac{\hbar^2 \Lambda^2}{2 m} - \mu(l) \right) \right]}
\label{2.13}
\end{eqnarray}

\begin{eqnarray}
F_{p-a} & = & F_{p-a} [\mu(l),T(l),\Gamma(l)] \nonumber \\
& = & \frac{1}{2 \Gamma(l) \left( \frac{\hbar^2 \Lambda^2}{2 m} - \mu(l) \right)} 
\coth \left[ \frac{\Gamma(l)}{2 k_B T(l)} \left( \frac{\hbar^2 \Lambda^2}{2 m} - \mu(l) \right) 
\right]
\label{2.14}
\end{eqnarray}

\begin{eqnarray}
F_{p-p} & = & F_{p-p} [\mu(l),T(l)] \nonumber \\
& = & \frac{1}{4 k_B T(l)} 
\frac{1}{\sinh^2 \left[ \frac{1}{2 k_B T(l)} \left( \frac{\hbar^2 \Lambda^2}{2 m} - \mu(l) 
\right) \right]}
\label{2.15}
\end{eqnarray}

\begin{eqnarray}
F_f & = & F_f [\mu(l),T(l),\Gamma(l)] \nonumber \\
& = & k_B T(l) \ln \left\{ 1 - \exp \left[ - \frac{\Gamma(l)}{k_B T(l)} \left( 
\frac{\hbar^2 \Lambda^2}{2 m} - \mu(l) \right) \right] \right\} \nonumber \\
& - & k_B T(l) \ln \left[ 1 - \exp \left( - \frac{\Gamma(l)}{k_B T(l)} \right) \right]
\label{2.16}
\end{eqnarray}
Following the Ref. \cite{12,13} we will solve these equations for $d<2$ ($\epsilon = 2-d$) and
for the case $d=2$ which is of special interest.

\section{Below two dimensions}

In order to solve the scaling equations (\ref{2.8}) - (\ref{2.12}) given in the previous section
we start with the equation (\ref{2.11}) for the coupling constant at $T=0$. Introducing the 
notation:
\begin{equation}
C = \frac{m K_d \Lambda^{d-2} \Gamma^2(l)}{2 \hbar^2}
\label{3.1}
\end{equation}
this equation becomes:
\begin{equation}
\frac{d u(l)}{d l} = \epsilon u(l) - C u^2(l)
\label{3.2}
\end{equation}
with the initial condition $u(l=0) = u_0$. The fixed point of Eq. (\ref{3.2}) has the simple 
form:
\begin{equation}
u^* = \frac{\epsilon}{C}
\label{3.3}
\end{equation}
and the general solution:
\begin{equation}
u(l) = \frac{u^*}{1 - \left( 1 - \frac{u^*}{u_0} \right) \exp \left( - \frac{\epsilon}{l} \right)}
\label{3.4}
\end{equation}
This solution satisfies the conditions:
\begin{equation}
u(l=0) = u_0
\label{3.5}
\end{equation}
\begin{equation}
u(l \rightarrow \infty) = u^*
\label{3.6}
\end{equation}
and in order to perform the calculation of physical quantities in the critical region we will use 
the linearized form of Eq. (\ref{3.4})
\begin{equation}
u(l) \simeq u^* + \frac{u^*}{u_0} (u_0  - u^*) e^{- \epsilon l} 
\label{3.7}
\end{equation}
In the limit of low temperatures the chemical potential $\mu(l)$ given by Eq. (\ref{2.10}) 
becomes:
\begin{equation}
\mu(l) = - K_d \lambda^d e^{2l} \int_0^l \; d l' \;
\frac{u(l') e^{-2l'}}{\exp \left[ \frac{\hbar^2 \Gamma(l') \lambda^2}{2 m k_B T(l')} e^{-2l'} 
- \frac{\Gamma(l') \mu(l')}{2 m k_B T(l')} \right] - 1}
\label{3.8}
\end{equation}
From Eq. (\ref{2.8}) and (\ref{2.9}) we obtain:
\begin{equation}
\Gamma(l) = \Gamma e^{-(2-d)l}
\label{3.9}
\end{equation}
\begin{equation}
T(l) = T e^{2l}
\label{3.10}
\end{equation}
and using for $\mu(l)$ the lowest approximation ($\mu(l) = \mu e^{2l}$) we calculate the second
term from the exponential of Eq. (\ref{3.8}) as:
\begin{equation}
\frac{\Gamma \mu}{2 m k_B T} \ll 1
\label{3.11}
\end{equation}
This inequality is satisfied even in the low temperatures domain, because $\Gamma^{-1}$ is the 
energy scale of the quantum fluctuations and in this domain these are important. Using these 
considerations, Eq. (\ref{3.8}) we get:
\begin{equation}
\mu(l) \simeq - K_d \Lambda^d e^{2l} \int_0^l \; d l' \; 
\frac{e^{-2l'} u(l')}{\exp \left( \frac{\hbar^2 \Gamma \Lambda^2}{2 m k_B T} \right) - 1} 
\label{3.12}
\end{equation}
and in order to calculate the condensed density $n$ and the coherence length $\xi$ we will use 
Eq. (\ref{3.7}) for $u(l)$. The renormalization procedure will be stoped for $l = l^*$, given by
the equation:
\begin{equation}
\mu(l^*) = - \alpha \frac{\hbar^2 \Lambda^2}{2 m}
\label{3.13}
\end{equation}
where $\alpha \lesssim 1$. If we introduce the notations:
\begin{equation}
a = \frac{\hbar^2 \Lambda^2 \Gamma}{2 m k_B T}
\label{3.14}
\end{equation}
\begin{equation}
b = 1 + \frac{\epsilon}{2}
\label{3.15}
\end{equation}
we obtain from Eqs. (\ref{3.12}) and (\ref{3.13})
\begin{equation}
e^{-2l^*} \left( 1 + \frac{2 \epsilon}{\alpha} \right) \simeq \frac{4 \epsilon}{\alpha} I(l^*)
\label{3.16}
\end{equation}
where
\begin{equation}
I(l^*) = \int_0^{l^*} \; d l' \; \frac{e^{-2bl'}}{\exp \left( a e^{-2l'} \right) - 1}
\label{3.17}
\end{equation}
This integral can be expressed as:
\begin{equation}
I(l^*) \simeq \frac{1}{2 a^b} \ln \frac{1}{A} 
\left[ 1 - \frac{\epsilon}{2} \frac{F(A)}{\ln \frac{1}{A}} \right]
\label{3.18}
\end{equation}
where $A = 2 \epsilon$ and
\begin{equation}
F(A) = \int_A^{\infty} \; dx \; x^{\frac{\epsilon}{2} - 1} \ln \left( 1 - e^{-x} \right)
\label{3.19}
\end{equation}
In the limit of low temperatures, Eq. (\ref{3.16}) gives:
\begin{equation}
e^{-2l'} = 2 \epsilon \left[ \frac{2 m k_B T \Gamma}{\hbar^2 \Lambda^2} \right]^{1 + \epsilon}
\ln \frac{1}{2 \epsilon}
\label{3.20}
\end{equation}
Using these result we can calculate the density of the condensed bosons as:
\begin{equation}
n = e^{-d l^*} \int \; \frac{d^d {\bf k}}{(2 \pi)^d} \; 
\frac{1}{\exp \left[ \frac{1}{k_B T(l^*)} \left( \frac{\hbar^2 k^2}{2m} 
- \mu(l^*) \right) \right] -1}
\label{3.21}
\end{equation}
The integral from Eq. (\ref{3.21}),
\begin{equation}
A = K_d \int_0^{\infty} \; d k \; 
\frac{k^{d-1}}{\exp \left[ \frac{1}{k_B T(l^*)} \left( \frac{\hbar^2 k^2}{2m} - \mu(l^*) \right) \right] -1}
\label{3.22}
\end{equation}
will be transformed using the notation
\begin{equation}
x = \frac{\hbar^2 k^2}{2 m k_B T(l^*)} - \frac{\mu(l^*)}{k_B T(l^*)}
\label{3.23}
\end{equation}
as
\begin{equation}
A = \frac{1}{2} K_d \left[ \frac{2m k_B T(l^*)}{\hbar^2} \right]^{(d/2)} 
\int_{|x_m|}^{\infty} \; d x \; \frac{x^{(d-2)/2}}{e^x -1} 
\label{3.24}
\end{equation}
where $|x_m| = \mu(l^*)/k_B T(l^*)$. In order to perform the integral given by Eq. (\ref{3.24})
we consider $\epsilon$ small ($d$ close to $d=2$) and in this case
\begin{equation}
A \simeq - \frac{1}{2} K_d \left[ \frac{2mk_B T(l^*)}{\hbar^2} \right]^{d/2}
\ln \left[ 1 - \exp \left( - \frac{\mu(l^*)}{k_B T(l^*)} \right) \right]
\label{3.25}
\end{equation}
Using $T(l^*) = T e^{2l^*}$ and Eq. (\ref{3.13}) we obtain for the condensed density
\begin{equation}
n(T) \simeq \frac{1}{4 \pi} \left[ \frac{2mk_B T}{\hbar^2} \right] 
\ln \frac{1}{2 \epsilon \ln \frac{1}{2 \epsilon}}
\label{3.26}
\end{equation}
The temperature dependence of the coherence length $\xi(T)$ will be calculated using
\begin{equation}
\xi^{-2}(T) = \frac{2m}{\hbar^2} |\mu(l^*)|
\label{3.27}
\end{equation}
where $\mu(l^*)$ is given by Eq. (\ref{3.13}). This gives
\begin{equation}
\xi(T) = \frac{1}{\Lambda \alpha^{1/2}} e^{l^*}
\label{3.28}
\end{equation}
and using Eq. (\ref{3.20})
\begin{equation}
\xi(T) = \frac{1}{\left( 2 \epsilon \ln \frac{1}{2 \epsilon} \right)^{1/2}} 
\frac{\hbar}{(2 m k_B)^{1/2}} T^{- \frac{1}{2} \left( 1 + \frac{\epsilon}{2} \right)}
\label{3.29}
\end{equation}
if we take $\alpha \simeq 1$.

This behavior appear as universal and is not surprising because as it was mentioned in 
Ref. \cite{13} the first contribution to the calculation of chemical potential in the fixed 
point gives a first term as universal.
 
\section{Two Dimensions}

The dilute Bose gas at $T=0$ was studied in Ref. \cite{13} and is known as marginal. In this
case we consider the system at finite temperature in the low temperatures approximation. The 
free energy and the specific heat will be also calculated for this system taking $z=2$. The high
temperature limit will be also analised.

\subsection{Low temperature limit}

\subsubsection{Scaling equations}

In this case the scaling equations have the form:
\begin{equation}
\frac{d \Gamma(l)}{d l} = 0
\label{4.1}
\end{equation}
\begin{equation}
\frac{d T(l)}{d l} = 2 T(l)
\label{4.2}
\end{equation}
\begin{equation}
\frac{d u(l)}{d l} = - \frac{m K_2}{2 \hbar^2} u^2(l)
\label{4.3}
\end{equation}
\begin{equation}
\frac{d \mu(l)}{d l} = 2 \mu(l) 
- \frac{\Lambda^2 K_2 u^2(l)}{\exp \left( \frac{\hbar^2 \Lambda^2}{2 m k_B T(l)} \right) -1}
\label{4.4}
\end{equation}
The Eqs. (\ref{4.1}) - (\ref{4.3}) have the solutions:
\begin{equation}
\Gamma(l) = \Gamma
\label{4.5}
\end{equation}
\begin{equation}
T(l) = T e^{2l}
\label{4.6}
\end{equation}
\begin{equation}
u(l) = \frac{4 \pi \hbar^2}{m} \; \frac{1}{l+l_0}
\label{4.7}
\end{equation}
where $\Gamma$ will be take as $\Gamma = 1$ and $l_0$ has been calculated as:
\begin{equation}
l_0 = \frac{4 \pi \hbar^2}{m u_0}
\label{4.8}
\end{equation}
The solution of Eq. (\ref{4.4}) has the form:
\begin{equation}
\mu(l) = - \frac{4 \Lambda^2 \hbar^2}{2 m} \int_0^l \; \frac{d l'}{l'+l_0} 
\; \frac{e^{-2l'}}{\exp \left( \frac{\hbar^2 \Lambda^2}{2 m k_B T} e^{-2l'} \right) -1}
\label{4.9}
\end{equation}
which will be written as:
\begin{eqnarray}
\mu(l) & = & - \frac{4 \Lambda^2 \hbar^2}{2 m} e^{2l} \; \frac{2 m k_B T}{\hbar^2 \Lambda^2}
\left\{ \frac{1}{2 l_0} \ln \left[ 1 - \exp \left( - \frac{\hbar^2 \Lambda^2}{2 m k_B T} \right)
\right] \right. \nonumber \\
& & \left. - \frac{1}{2 l_0} \left( 1 + \frac{l}{l_0} \right)^{-1} 
\ln \left[ 1 - \exp \left( - \frac{\hbar^2 \Lambda^2}{2 m k_B T} e^{-2l} \right) \right] 
\right\} - \frac{2 m k_B T}{\hbar^2 \Lambda^2} \; F(l)
\label{4.10}
\end{eqnarray}
where
\begin{equation}
F(l) = \int_0^{2l} \; \frac{d x}{(x+2l_0)^2} \; 
\ln \left[ 1 - \exp \left( - \frac{\hbar^2 \Lambda^2}{2 m k_B T} e^{-x} \right) \right]
\label{4.11}
\end{equation}
The renormalization procedure will be stoped at $l^*$ defined also by:
\begin{displaymath}
\mu(l^*) = - \alpha \frac{\hbar^2 \Lambda^2}{2 m}
\end{displaymath}
but in this case $\mu(l)$ is given by Eq. (\ref{4.10}).
Following the same procedure we calculate
\begin{equation}
e^{-2l^*}  = \frac{4}{\alpha} \left[ \frac{4}{\alpha} - \ln \frac{4}{\alpha} \right]
\; \frac{2 m k_B T}{\hbar^2 \Lambda^2} \; 
\frac{1}{\ln \frac{\alpha}{4} \frac{\hbar^2 \Lambda^2}{2 m k_B T}}
\label{4.12}
\end{equation}
and if we introduce the effective temperature $T_0 = \hbar^2 \Lambda^2 \alpha /8 m k_B$ 
Eq. (\ref{4.12}) will be written as:
\begin{equation}
e^{-2l^*} = \frac{C(\alpha)}{4} \; \frac{T}{T_0} \; \frac{1}{\ln \frac{T_0}{T}}
\label{4.13}
\end{equation}
Using Eq. (\ref{3.28}) we get for coherence length
\begin{equation}
\xi (T) \sim \left| \frac{\ln \frac{T_0}{T}}{\frac{T}{T_0}} \right|^{1/2}
\label{4.14}
\end{equation}
The number of condensate bosons has been calculated using a similar relation with (\ref{3.21}) 
and we get:
\begin{equation}
n(T) = \frac{2 m k_B T}{4 \pi \hbar^2} \; 
\ln \frac{1}{1 - \exp \left( - \frac{\hbar^2 \Lambda^2}{2 m k_B T} e^{-2l^*} \right)}
\label{4.15}
\end{equation}
which gives
\begin{equation}
n(T) = \frac{2 m k_B T}{4 \pi \hbar^2} \; \ln \ln \left( \frac{T_0}{T} \right)
\label{4.16}
\end{equation}

\subsubsection{The free energy and specific heat}

The free energy given by general Eq. (\ref{2.1}) has the form:
\begin{equation}
\frac{d F(l)}{d l} = 4 F(l) + C_f (\mu(l),T(l))
\label{4.17}
\end{equation}
with 
\begin{equation}
C_f = K_2 \Lambda^2 k_B T \ln \left\{ 1 - \exp \left[ - \frac{\Gamma(l)}{k_B T}
\left( \frac{\hbar^2 \Lambda^2}{2 m} - \mu(l) \right) \right] \right\}
\label{4.18}
\end{equation}
Using the substitution $T = T e^{2x}$ we write the solution of Eq. (\ref{4.17}) as:
\begin{equation}
F(l) = \int_0^l \; d x \; e^{-4x} C_f \left( T e^{2x} \right)
\label{4.19}
\end{equation}
where we approximate $C_f$ as:
\begin{equation}
C_f = K_2 \Lambda^2 k_B T 
\ln \left[ 1 - \exp \left( - \frac{\Gamma(l)}{k_B T} - \mu(l) \right) \right]
\label{4.20}
\end{equation}
The expression for $F(l^*)$ becomes:
\begin{equation}
F(l^*) = \frac{\Lambda^2}{2 \pi} k_B T \; \int_0^{l^*} \; dx \; e^{-2x} 
\ln \left( 1 - e^{- \frac{A}{T} e^{-2x}} \right)
\label{4.21}
\end{equation}
where 
\begin{displaymath}
A = \frac{\Gamma}{k_B} \left( \frac{\hbar^2 \Lambda^2}{2 m} - \mu \right)
\end{displaymath}
In the low temperature limit Eq. (4.21) will be approximated as:
\begin{eqnarray}
F & = & \frac{\Lambda^2}{4 \pi} k_B T \ln \frac{A}{T} 
\left[ 1 - \frac{\frac{T}{T_0} \frac{C(\alpha)}{4}}{\ln \frac{T_0}{T}} \right] \nonumber \\
& + & \frac{\Lambda^2}{4 \pi} (k_B T)^2 \; 
\frac{\frac{T}{T_0}}{\ln \left| \frac{T}{T_0} \right|} \; 
\ln \left[ \frac{\ln \left| \frac{T}{T_0} \right|}{\left| \frac{T}{T_0} \right|} \right] \nonumber \\
& + & \frac{\Lambda^2}{4 \pi} k_B T \; 
\left[ \frac{\frac{T}{T_0}}{\ln \left| \frac{T}{T_0} \right|} - 1 \right]
\label{4.22}
\end{eqnarray}
From this equation we calculate the specific heat $C_v(T) = - T \partial^2 T / \partial T^2$ and
the dominant contribution in temperature has the form:
\begin{equation}
C_v(T) = C_0 \frac{\left| \frac{T}{T_0} \right|}{\ln^3 \left| \frac{T}{T_0} \right|}
\label{4.23}
\end{equation}
where $C_0 = C_0(\Lambda)$ which shows that the result is $\Lambda$ - dependent.

\subsection{High temperature limit}

The high temperature domain is called "classical" domain, because it is dominated by the classical 
fluctuations of the $\phi({\bf x},\tau)$ and is usually defined by $z=0$.

\subsubsection{Scaling equations}

From the general equations (\ref{2.8}) - (\ref{2.11}) we write these equations in the high 
temperature limit as: 
\begin{equation}
\frac{d \left[ T(l) \Gamma^{-1}(l) \right]}{d l} = 2 \left[ T(l) \Gamma^{-1}(l) \right]
\label{5.1}
\end{equation}
\begin{equation}
\frac{d \mu(l)}{d l} = 2 \mu(l) - \frac{1}{2 \pi} {\tilde F}_{\mu} [\mu(l),T(l)] v(l)
\label{5.2}
\end{equation}
\begin{equation}
\frac{d v(l)}{d l} = 2 v(l) - \frac{5}{4 \pi} {\tilde F}_v [\mu(l),T(l)] v^2 (l)
\label{5.3}
\end{equation}
where
\begin{equation}
v(l) = k_B T u(l)
\label{5.4}
\end{equation}
and ${\tilde F}_{\mu}$, ${\tilde F}_v$ have been obtained from Eqs. (\ref{2.13}) - (\ref{2.15})
in the limit $T \rightarrow \infty$ as:
\begin{equation}
{\tilde F}_{\mu} [\mu(l),T(l)] \simeq 
\frac{\Lambda^2}{\left[ \frac{\hbar^2 \Lambda^2}{2 m} - \mu(l) \right]} 
\label{5.5}
\end{equation}
\begin{equation}
{\tilde F}_v [\mu(l),T(l)] \simeq 
\frac{\Lambda^2}{\left[ \frac{\hbar^2 \Lambda^2}{2 m} - \mu(l) \right]^2}
\label{5.6}
\end{equation}
In order to solve the Eqs. (\ref{5.1}) - (\ref{5.3}) we have to define ${\tilde l}$, the value 
of $l$ at which the flow enters in the classical regime defined by
\begin{equation}
\frac{T(l)}{\Gamma(l)} \gg 1
\label{5.7}
\end{equation}
and introduce
\begin{equation}
\mu({\tilde l}) = {\tilde \mu}_0 \; , \; u({\tilde l}) = {\tilde u}_0 \; , \; 
v({\tilde l}) = {\tilde v}_0
\label{5.8}
\end{equation}
In order to simplify the calculation we perform a simple transformation $l' = l - {\tilde l}$ 
which make the flow to start at $l' = 0$. 

The new scaling equations describing the classical regime will be:
\begin{equation}
\frac{d [T(l') \Gamma^{-1}(l')]}{d l'} = 2 [T(l') \Gamma^{-1}(l')]
\label{5.9}
\end{equation}
\begin{equation}
\frac{d \mu(l')}{d l'} = 2 \mu(l') - \frac{1}{2 \pi} 
\frac{\Lambda^2 v(l')}{\frac{\hbar^2 \Lambda^2}{2 m} - \mu(l')}
\label{5.10}
\end{equation}
\begin{equation}
\frac{d v(l')}{d l'} \simeq 2 v(l') - \frac{5 m^2}{\pi \hbar^4} \frac{v^2 (l')}{\Lambda^2} 
\label{5.11}
\end{equation}
The equation (\ref{5.11}) has been obtained neglecting $\mu(l')$ in Eq. (\ref{5.6}). In this
form the equation can be solved and we obtain the exact solution
\begin{equation}
v(l') = \frac{2 {\tilde v}_0}{B {\tilde v}_0 + (2 - B {\tilde v}_0) \exp{(-2l')}}
\label{5.12}
\end{equation}
where $B = 5 m^2 / \pi \hbar^4 \Lambda^2$.

We define ${l'}_*$ a value of $l'$ for which we stop the scaling by a similar condition as in the
low temperature case,
\begin{equation}
v({l'}_*) = 1
\label{5.13}
\end{equation}
which gives
\begin{equation}
\exp{(2 {l'}_*)} = \frac{2 - B {\tilde v}_0}{B {\tilde v}_0 - 2 {\tilde v}_0}
\label{5.14}
\end{equation}
and from this equation we calculate
\begin{equation}
{l'}_* \simeq \frac{1}{2} \ln \frac{1}{{\tilde v}_0}
\label{5.15}
\end{equation}
where we used $u T({l'}_*) = 1$.

The Eq. (\ref{5.10}) for the chemical potential can also be solved using for $v(l')$ the 
expression (\ref{5.12}) and we get:
\begin{equation}
\mu(l') = e^{2 l'} \left[ {\tilde \mu}_0 - \frac{2 m}{\pi \hbar^2 B} l'
- \frac{m}{\pi \hbar^2 B} \ln \left( e^{-2 l'} 
+ \frac{B {\tilde v}_0}{2 - B {\tilde v}_0} \right) \right]
\label{5.16}
\end{equation}
where ${\tilde \mu}_0$ is given by:
\begin{equation}
{\tilde \mu}_0 = \frac{m}{\pi \hbar^2 B} {l'}_* + \frac{m}{\pi \hbar^2 B}
\ln \left( e^{-2 {l'}_*} + \frac{B {\tilde v}_0}{2 - B {\tilde v}_0} \right)
\label{5.17}
\end{equation}
Using this equation $\mu(l')$ is approximated as:
\begin{equation}
\mu(l') \simeq e^{2 l'} \left[ \frac{2 m {\tilde v}_0}{\pi \hbar^2} ({l'}_* - l') \right]
\label{5.18}
\end{equation}

\subsubsection{Density of bosons}

In order to calculate the bosonic density, we will use the general equations:
\begin{equation}
n = - \frac{\partial F}{\partial \mu}
\label{5.19}
\end{equation}
where the free energy will be written as:
\begin{equation}
F(l') = \frac{1}{2 \pi} \int_0^{l'} \; d x \; e^{-2x} T(x) 
\ln \left\{ 1 - \exp \left[ - \frac{\Gamma(x)}{k_B T(x)} 
\left( \frac{\hbar^2 \lambda^2}{2 m} - \mu(x) \right) \right] \right\}
\label{5.20}
\end{equation}
Using the relations:
\begin{equation}
n = - \int_0^{{l'}_*} \; dl' \; \frac{\partial}{\partial \mu(l')} 
\left( \frac{\partial F}{\partial l'} \right) \frac{d \mu(l')}{d \mu} 
\label{5.21}
\end{equation}
and for $\mu(l')$ the simple approximation
\begin{equation}
\mu(l') = \mu({\tilde l}) e^{2(l' + {\tilde l})}
\label{5.22}
\end{equation}
where $\mu({\tilde l}) \simeq \mu$, the Eq. (\ref{5.21}) becomes:
\begin{equation}
n = - \int_0^{{l'}_*} \; dl' \; e^{2(l'+{\tilde l})} \; 
\frac{\partial}{\partial \mu(l')} \left( \frac{\partial F(l')}{\partial l'} \right)
\label{5.23}
\end{equation}
From Eqs. (\ref{5.20}) and (\ref{5.23}) we calculate the general equation for bosonic density
$n$ as:
\begin{equation}
\frac{n}{T} = \frac{1}{2 \pi} \int_0^{{l'}_*} \; dl_1 \; 
\frac{\Gamma(l_1) / k_B T(l_1)}{\exp \left\{ \Gamma(l_1) / k_B T(l_1) 
\left[ \frac{\hbar^2 \Lambda^2}{2 m} - \mu(l_1) \right] \right\} - 1}
\label{5.24}
\end{equation}
This integral will be approximated in the high temperature domain as:
\begin{equation}
\frac{n}{T} \simeq \frac{1}{2 \pi} \int_0^{{l'}_*} \; d l_1 \; \frac{2 m}{\hbar^2 \Lambda^2}
\left[ 1 + \frac{\hbar^2 \Lambda^2}{2 m} \mu(l_1) + \cdots \right] \; \simeq \; 
\frac{1}{2 \pi} \frac{2 m}{\hbar^2 \Lambda^2} {l'}_*
\label{5.25}
\end{equation}
and from Eq. (\ref{5.15}) we calculate:
\begin{equation}
\frac{n}{T} \simeq \frac{m}{\pi \hbar^2} \ln \frac{1}{{\tilde u}_0}
\label{5.26}
\end{equation}
which is the density of bosons in the classical regime. This result gives us the possibility to
perform a matching with the low temperatures regime. Indeed, using 
${\tilde u}_0 \sim 1 / \ln T_0$ for $T_0 \ll 1$ we reobtain the expression for
$n_c \sim T \ln \ln 1/T$ the well-known result from Ref. \cite{5}.

\section{Discussions}

We reconsidered the problem of dilute Bose gas at finite temperature using the RG method in 
$d=2$. This problem has been considered first in the well-known paper by Fisher and Hohenberg
\cite{12}, but their calculation may give divergencies in the integrals contained in the 
physical quantities calculated in the quantum regime, because of an approximation for the 
coupling constant which is taken of the form $v(l) \sim 1/l$.

We solved the problem by using a linearization of the exact solution near the fixed point for
$d<2$ and taking the exact solution in $d=2$. We also calculate the coherence length $\xi(T)$
and the specific heat from the free energy. As a common point of view with the classical Fisher
and Hohenberg paper we mention the non - universal behavior of the system. The new features which
appear can be concluded as follows for $d<2$ :
\begin{itemize}

\item the bosonic density $n(T)$ present the specific $T^{d/2}$ dependence as well as 
$\ln (1/\epsilon \ln 1/\epsilon )$ dependence;

\item the coherence length $\xi(T)$ calculated also has 
$1/ \left[ T^{\frac{1}{2}(1 + \frac{\epsilon}{2})} \epsilon \ln 1/\epsilon \right]^{1/2}$
dependence;

\end{itemize}

For the two dimensional case $d=2$ we used the exact solution for the coupling constant
$u(l) \sim 1/ (l+l_0)$ and we showed that :

\begin{itemize}

\item the density $n(T)$ is non - universal and $n(T) \sim T \ln 1/T$, which gives for the 
critical temperature the Fisher and Hohenberg result, $\ln \ln 1/na^2$, where 
$\Lambda^2 = 1/a^2$, $a$ being the potential range;

\item the coherence length $\xi(T)$ is non - universal and depends of $T$ as 
$\xi(T) \sim |(\ln T)/ T|^{1/2}$;

\item the specific heat $C(T)$ is also non - universal and $C_v(T) \sim T/ |\ln T|^3$;

\end{itemize}

The RG scaling equation have been used also in the high temperature domain for $d=2$. The main
result of this treatment is the calculation of the bosonic density which can be matched with the
result from the low temperature regime by choosing the coupling constant as the solution of the
scaling equations from the low temperature regime.

\end{document}